# Deformation effects in the alpha accompanied cold ternary fission of even-even $^{244-260}$Cf isotopes

K. P. Santhosh* and Sreejith Krishnan

School of Pure and Applied Physics, Kannur University, Swami Anandatheertha Campus, Payyanur 670327, Kerala, INDIA
email: drkpsanthosh@gmail.com

**Abstract.**

Within the Unified ternary fission model (UTFM), the alpha accompanied ternary fission of even-even $^{244-260}$Cf isotopes has been studied by taking the interacting barrier as the sum of Coulomb and proximity potential. For the alpha accompanied ternary fission of $^{244}$Cf isotope, the highest yield is obtained for the fragment combination $^{108}$Ru+$^{4}$He+$^{132}$Te, which contain near doubly magic nuclei $^{132}$Te (N=80, Z=52). In the case of $^{246}$Cf and $^{248}$Cf isotopes, the highest yield is obtained for the fragment combinations with near doubly magic nuclei $^{134}$Te (N=82, Z=52) as the heavier fragment. The highest yield obtained for $^{250}$Cf, $^{252}$Cf, $^{254}$Cf, $^{256}$Cf, $^{258}$Cf and $^{260}$Cf isotopes is for the fragment combination with doubly magic nuclei $^{132}$Sn (N=82, Z=50) as the heavier fragment. We have included the effect of deformation and orientation of fragments and this has revealed that in addition to closed shell effect, ground state deformation also plays an important role in the calculation of relative yield of favorable fragment combinations. The computed isotopic yields for alpha accompanied ternary fission of $^{252}$Cf isotope are found to be in agreement with the experimental data. The emission probability and kinetic energy of long range alpha particle is calculated for the various isotopes of Cf and are found to be in good agreement with the experimental data.

## 1. Introduction

The light charged particle accompanied fission also termed as ternary fission is a fission process in which three fragments are formed and it has been an interesting area for discussion under both experimental and theoretical efforts for the last few decades [1-8]. The cold alpha accompanied ternary fragmentation of $^{248}$Cm isotope was studied by Sandulescu et al. [9] using double folding potential generated by Coulomb plus M3Y nucleon-nucleon forces. The energy released from the cold ternary fission of heavy and superheavy nuclei with Z=90-116 was calculated by Poenaru et al. [10], in which the $Q$ values were calculated for the cold ternary fragmentation of nearly equal sized fragments and also for the different light charged particle accompanied fission processes. Carstoiu et al. [11] evaluated the lifetime of trinuclear molecules with two deformed fragments and an alpha particle, and the lifetimes of such trinuclear molecules were found to be of the order of one second. The dynamics for the distribution of ternary cluster preformation probability for the alpha and $^{10}$Be accompanied ternary fission of $^{252}$Cf isotope along the fission trajectory were studied by Florescu et al. [12].

Experimentally, the energy spectra of long range alpha particles emitted from the spontaneous ternary fission of $^{250}$Cf, $^{256}$Fm and $^{257}$Fm isotopes were measured by Wild et al. [13] and they also determined the emission probabilities for the triton, proton and alpha particle. Ralph [14] measured the probability for the emission of long range alpha particles from the spontaneous fission of $^{252}$Cf, $^{242,244}$Cm and $^{240,242}$Pu isotopes and came to the conclusion that the emission probability of long range alpha particles increased with increasing $Z^2/A$ values. The alpha accompanied ternary fission of $^{252}$Cf isotope was observed by Ramayya et al. [15,16] using triple gamma coincidence technique with Gammasphere with 72 gamma-ray detectors and the highest yield was obtained for the Zr-Ba fragment combination. The yields of both ternary and binary spontaneous fission of $^{252}$Cf isotope were found out and the results were compared with the theoretical calculations using M3Y nucleon-nucleon potential and WKB approximation for barrier penetration. In addition to alpha particle, $^{10}$Be and $^{14}$C accompanied spontaneous ternary fission was also studied by Hamilton et al. [17-19] using the Gammasphere technique for $^{252}$Cf isotope. Based on the theoretical model of Carjan [20], the characteristics of long range alpha particles were studied in the spontaneous ternary fission of $^{244}$Cm, $^{246}$Cm, $^{248}$Cm, $^{250}$Cf, $^{252}$Cf and even-even $^{238-244}$Pu isotopes [21-23]. The emission probability and kinetic energy distribution of long range alpha particles were studied in detail for the corresponding isotopes of Cm, Pu and Cf, with the suitable ΔE-E telescope detectors at IRMM.

A new type of decay mode called collinear cluster tripartition (CCT) was introduced by Pyatkov et al. [24] for the spontaneous ternary fission of $^{252}$Cf isotope and the studies [25,26] based on the collinear cluster tripartition of $^{236}$U and $^{252}$Cf isotopes revealed that the isotopes of Si and Ca respectively connect the isotopes of Sn to the edge fragment $^{68}$Ni. Recently, Vijayaraghavan et al. [27,28] calculated the ternary potential energy surfaces for the fragment combinations by changing the angle between the end fragments with respect to the middle fragment and hence analyzed the role of arrangements of fragments in the ternary fission process of $^{252}$Cf isotope. Based on the three cluster model (TCM) proposed by Manimaran et al. [29-31], the authors had also studied the possibilities of the ternary fragmentation into three nearly equal sized fragments in the case of $^{252}$Cf isotope and came to the conclusion that the true ternary fission was energetically possible only for the fragment combinations with minimum fragmentation potential and high Q value. For the first time, the effects of deformation and orientation of fragments were included by Manimaran et al. [32] in the three-cluster model for the study of $^4$He and $^{10}$Be accompanied ternary fission of $^{252}$Cf isotope.

Using our recently proposed Unified ternary fission model (UTFM), we have done a number of theoretical calculations and predictions on the ternary fission of various isotopes of Cm, Cf and Pu. We have calculated the relative yield for $^{242}$Cm isotope with $^4$He, $^{10}$Be and $^{14}$C as light charged particle and the long range alpha emission probability for the various ternary fragmentations of $^{242}$Cm using UTFM [33]. With $^{34}$Si as light charged particle, the ternary fission of $^{242}$Cm [34] has been studied in equatorial and collinear configurations. Moreover, in the case of $^{252}$Cf isotope [35], the relative yield has been calculated for the alpha accompanied ternary fission and the obtained results have been compared with the experimental data. A comparative study has also been done for the relative yield between the alpha accompanied ternary fission and binary fission of even-even $^{244-252}$Cm isotopes [36]. The kinetic energy of alpha particle, probability for the emission of long range alpha particle and the effect of deformation and orientation of fragments have also been studied in detail for even-even $^{244-252}$Cm [36] and $^{238-244}$Pu [37] isotopes. From our earlier works on ternary fission with alpha particle as the middle fragment, we arrived at the experimentally proved fact that the equatorial emission of fragments is the most favorable configuration than the collinear emission of fragments. Also, it is established that, the fragment combinations which possess highest yields are found to be the same in both equatorial and collinear configurations when the magnitude of relative yield is not considered. This is because the initial phase for both equatorial and collinear configurations is found to be the same as the size of alpha particle is small compared to the main fission fragments. Hence in the present work, we have calculated the driving potential and relative yield for the ternary fission of even-even $^{244-260}$Cf isotopes with $^4$He as light charged particle in equatorial configuration. As an extension of our earlier works, the kinetic energy of long range alpha particle, emission probability of long range alpha particle and the effect of deformation and orientation of fragments are also studied in detail in the alpha accompanied ternary fragmentation of even-even $^{244-260}$Cf isotopes.

The formalism used for our calculation is described in Section 2. The results and discussions on the alpha accompanied ternary fission of even-even $^{244-260}$Cf isotopes are given in Section 3 and we summarize the entire work in Section 4.

## 2. Unified Ternary Fission Model (UTFM)

The light charged particle accompanied ternary fission is energetically possible only if $Q$ value of the reaction is positive. ie.

$$Q = M - \sum_{i=1}^{3} m_i > 0 \tag{1}$$

Here $M$ is the mass excess of the parent and $m_i$ is the mass excess of the fragments. The interacting potential barrier for a parent nucleus exhibiting cold ternary fission consists of Coulomb potential and nuclear proximity potential of Blocki et al [38, 39]. The proximity potential was first used by Shi and Swiatecki [40] in an empirical manner and has been quite extensively used by Gupta et al., [41] in the preformed cluster model (PCM) and is based on pocket formula of Blocki et al [38]. But in the present manuscript, another formulation of proximity potential (eqn (21a) and eqn (21b) of Ref. [39]) is used as given by Eqs. 7 and 8. The interacting potential barrier is given by,

$$V = \sum_{i=1}^{3} \sum_{j>i}^{3} (V_{Cij} + V_{Pij}) \tag{2}$$

with $V_{Cij} = \dfrac{Z_i Z_j e^2}{r_{ij}}$, the Coulomb interaction between the fragments. Here $Z_i$ and $Z_j$ are the atomic numbers of the fragments and $r_{ij}$ is the distance between fragment centres. The nuclear proximity potential [38] between the fragments is,

$$V_{Pij}(z) = 4\pi\gamma b \left[ \frac{C_i C_j}{(C_i + C_j)} \right] \Phi\left(\frac{z}{b}\right) \tag{3}$$

Here $\Phi$ is the universal proximity potential and $z$ is the distance between the near surfaces of the fragments. The distance between the near surfaces of the fragments for equatorial configuration is considered as $z_{12} = z_{23} = z_{13} = z$. The Süssmann central radii $C_i$ of the fragments related to sharp radii $R_i$ is,

$$C_i = R_i - \left(\frac{b^2}{R_i}\right) \tag{4}$$

For $R_i$ we use semi empirical formula in terms of mass number $A_i$ as [38]

$$R_i = 1.28A_i^{1/3} - 0.76 + 0.8A_i^{-1/3} \qquad (5)$$

The nuclear surface tension coefficient called Lysekil mass formula [42] is,

$$\gamma = 0.9517[1 - 1.7826(N-Z)^2/A^2] \text{ MeV/fm}^2 \qquad (6)$$

where $N$, $Z$ and $A$ represents neutron, proton and mass number of the parent, $\Phi$, the universal proximity potential (eqn (21a) and eqn (21b) of Ref. [39]) is given as,

$$\Phi(\varepsilon) = -4.41e^{-\varepsilon/0.7176} \text{ , for } \varepsilon > 1.9475 \qquad (7)$$

$$\Phi(\varepsilon) = -1.7817 + 0.9270\varepsilon + 0.0169\varepsilon^2 - 0.05148\varepsilon^3 \text{ , for } 0 \leq \varepsilon \leq 1.9475 \qquad (8)$$

with $\varepsilon = z/b$, where the width (diffuseness) of the nuclear surface $b \approx 1$ fermi.

Using one-dimensional WKB approximation, barrier penetrability $P$, the probability for the ternary fragments to cross the three body potential barrier is given as,

$$P = \exp\left\{-\frac{2}{\hbar}\int_{z_1}^{z_2}\sqrt{2\mu(V-Q)}dz\right\} \qquad (9)$$

The turning point $z_1 = 0$ represents touching configuration and $z_2$ is determined from the equation $V(z_2) = Q$, where $Q$ is the decay energy. The potential $V$ in eqn. 9, which is the sum of Coulomb and proximity potential is given by eqn. 2, and are computed by varying the distance between the near surfaces of the fragments. In eqn. 9 the mass parameter is replaced by reduced mass $\mu$ and is defined as,

$$\mu = m\left(\frac{\mu_{12}A_3}{\mu_{12} + A_3}\right) \qquad (10)$$

and

$$\mu_{12} = \frac{A_1 A_2}{A_1 + A_2} \qquad (11)$$

where $m$ is the nucleon mass and $A_1$, $A_2$ and $A_3$ are the mass numbers of the three fragments. The relative yield can be calculated as the ratio between the penetration probability of a given fragmentation over the sum of penetration probabilities of all possible fragmentations and is as follows,

$$Y(A_i, Z_i) = \frac{P(A_i, Z_i)}{\sum P(A_i, Z_i)} \qquad (12)$$

## 3. Results And Discussion

The role of closed shell effect or nearly closed shell effects arises from the cold reaction valley, which is justified through the Quantum mechanical fragmentation theory (QMFT) [43]. Using the concept of cold reaction valley, which was introduced in relation to the structure of minima in the so called driving potential, the alpha accompanied cold ternary fission of even-even $^{244-260}$Cf isotopes are studied in equatorial configuration. The driving potential is defined as the difference between the interaction potential $V$ and the decay energy $Q$ of the reaction, where the interaction potential $V$ is taken as the sum of Coulomb potential and nuclear proximity potential. The $Q$ values are calculated using the recent mass tables of Wang et al. [44] and for those isotopes for which the experimental values are not available; we have taken the values from the mass tables of Moller et al. [45]. For a fixed light charged particle $A_3$ attached to main fission fragments, the driving potential for a parent nucleus can be calculated for all possible fragment combinations as a function of mass and charge asymmetries respectively given as $\eta = \dfrac{A_1 - A_2}{A_1 + A_2}$ and $\eta_Z = \dfrac{Z_1 - Z_2}{Z_1 + Z_2}$. A pair of charges is singled out for every fixed mass pair $(A_1, A_2)$ in which the driving potential is minimized. The fragment combinations that occur in the cold reaction valley with a low driving potential and high Q value possess the highest yield in the ternary fission process. The schematic diagram for the equatorial emission of three spherical fragments at the touching configuration is shown in figure 1.

### 3.1. Alpha accompanied cold ternary fission of $^{244-260}$Cf isotopes

The driving potential is calculated for the alpha accompanied cold ternary fission of $^{244}$Cf isotope and is plotted as a function of fragment mass number $A_1$ as shown in figure 2. The minima found in the cold valley are at $^{4}$He, $^{6}$He, $^{10}$Be, $^{14}$C, $^{32}$Si, $^{34}$Si, $^{46}$Ar, $^{50}$Ca, $^{52}$Ca, $^{80}$Ge, $^{84}$Se etc. In the cold reaction valley plot, the fragment combination $^{4}$He+$^{4}$He+$^{236}$Pu possesses the least driving potential. The fragment combination around $^{108}$Ru+$^{4}$He+$^{132}$Te may possess the highest yield due to the presence of doubly or near doubly magic Te isotopes and also due to high Q value. Next the barrier penetrability is calculated for the alpha accompanied cold ternary fission of $^{244}$Cf isotope using the formalism described above. Using equation (12) relative yield is calculated and plotted as a function of fragment mass number $A_1$ and $A_2$ as shown in figure 3(a). From the figure it is clear that the combination $^{108}$Ru+$^{132}$Te+$^{4}$He has the highest yield due to the presence of near doubly magic nuclei $^{132}$Te (Z=52, N=80). The next highest yield can be observed for the fragment combinations $^{106}$Ru+$^{134}$Te+$^{4}$He and $^{104}$Mo+$^{136}$Xe+$^{4}$He which is due to the presence of near doubly magic nuclei $^{134}$Te (Z=52, N=82) and neutron shell closure N=82 of $^{136}$Xe respectively.

In the alpha accompanied ternary fission of $^{246-260}$Cf isotopes, the driving potential were calculated and studied as a function of fragment mass number $A_1$. In the cold reaction valley plots, the minima occur at $^{4}$He, $^{6}$He, $^{10}$Be, $^{14}$C, $^{16}$C, $^{22}$O, $^{24}$Ne, $^{28}$Mg, $^{32}$Si, $^{38}$S, $^{40}$S, $^{46}$Ar, $^{50}$Ca, $^{52}$Ca etc. The barrier penetrability and relative yield is calculated for the alpha accompanied ternary fission of $^{246}$Cf isotope and is plotted as a function of fragment mass number $A_1$ and $A_2$ as shown in figure 3(b). From the figure it is clear that the fragment combination $^{108}$Ru+$^{4}$He+$^{134}$Te has the highest yield which is due to the presence of nearly doubly magic nuclei $^{134}$Te (Z=52, N=82). The next highest yield can be observed for the fragment combination $^{114}$Pd+$^{4}$He+$^{128}$Sn which possess proton shell closure Z=50 of $^{128}$Sn. The other fragment combinations which possess highest yield are $^{112}$Pd+$^{4}$He+$^{130}$Sn and $^{110}$Ru+$^{4}$He+$^{132}$Te which is due to the presence of near doubly magic nuclei $^{130}$Sn (N=80, Z=50) and $^{132}$Te (N=80, Z=52) respectively.

The most possible fragment combination occur in the ternary fission of $^{248}$Cf is found by calculating the relative yield and plotted as a function of mass numbers $A_1$ and $A_2$ as shown in figure 3(c). The fragment combination $^{110}$Ru+$^{4}$He+$^{134}$Te has highest yield because of the near doubly magic nuclei $^{134}$Te (N=82, Z=52). The next highest yields are obtained for the fragment combinations $^{114}$Pd+$^{4}$He+$^{130}$Sn and $^{116}$Pd+$^{4}$He+$^{128}$Sn which possess near doubly magic nuclei $^{130}$Sn (N=80, Z=50) and proton shell closure Z=50 of $^{128}$Sn respectively.

In the case of $^{250}$Cf isotope, the barrier penetrability is calculated for each charge minimized fragments found in the cold valley plot and thereafter the relative yield is calculated and plotted as a function of fragment mass numbers $A_1$ and $A_2$ as shown in figure 3(d). The highest yield is found for the fragment combination $^{114}$Pd+$^4$He+$^{132}$Sn which is due to the doubly magic nuclei $^{132}$Sn (N=82, Z=50). The next highest yield is found for the fragment combination $^{116}$Pd+$^4$He+$^{130}$Sn which is due to nearly doubly magic nuclei $^{130}$Sn (N=80, Z=50). The yield obtained for the fragment combination $^{112}$Ru+$^4$He+$^{134}$Te is due to the presence of nearly doubly magic $^{134}$Te (N=82, Z=52).

In the case of $^{252}$Cf isotope, the relative yield is calculated for each charge minimized fragment combinations found in the cold valley plot and plotted as a function of fragment mass numbers $A_1$ and $A_2$ as shown in figure 3(e). The highest yield is found for the fragment combination $^{116}$Pd+$^4$He+$^{132}$Sn, in which $^{132}$Sn is a doubly magic nuclei (N=82, Z=50). The next highest yield is obtained for the fragment combinations $^{118}$Pd+$^4$He+$^{130}$Sn and $^{114}$Ru+$^4$He+$^{134}$Te which possess the presence of near doubly magic nuclei $^{130}$Sn (N=80, Z=50) and $^{134}$Te (N=82, Z=52) respectively.

Keeping $^4$He as the light charged particle, the relative yield is calculated for each charge minimized fragments found in the cold reaction valley of $^{254}$Cf isotope and plotted as a function of fragment mass numbers $A_1$ and $A_2$ as shown in figure 3(f). The highest yield is obtained for the fragment combination $^{118}$Pd+$^4$He+$^{132}$Sn, which possess high Q value and doubly magic nuclei $^{132}$Sn (N=82, Z=50). The next highest yield is obtained for the fragment combination $^{120}$Pd+$^4$He+$^{130}$Sn, which possess near doubly magic nuclei $^{130}$Sn (N=80, Z=50).

In the case of $^{256}$Cf isotope as the parent nucleus with $^4$He as the light charged particle (LCP), the relative yield is calculated and plotted as a function of fragment mass numbers $A_1$ and $A_2$ as shown in figure 3(g). The highest yield is obtained for the fragment combination $^{120}$Pd+$^4$He+$^{132}$Sn, which is due to the presence of doubly magic nuclei $^{132}$Sn (N=82, Z=50). The next highest yield is obtained for the fragment combination $^{122}$Pd+$^4$He+$^{130}$Sn, which possess near doubly magic nuclei $^{130}$Sn (N=80, Z=50).

The relative yield is calculated for alpha accompanied ternary fission of $^{258}$Cf isotope and plotted as a function of fragment mass numbers $A_1$ and $A_2$ as shown in figure 3(h). The highest yield is found for the fragment combination $^{122}$Pd+$^4$He+$^{132}$Sn, which possess doubly magic nuclei $^{132}$Sn (N=82, Z=50) and the next highest yield is obtained for the fragment combination $^{126}$Cd + $^{128}$Cd + $^4$He.

For $^{260}$Cf isotope the relative yield is calculated and plotted as a function of fragment mass numbers $A_1$ and $A_2$ as shown in figure 3(i). The highest yield is found for the fragment combination $^{124}$Pd+$^4$He+$^{132}$Sn, which possess doubly magic nuclei $^{132}$Sn (N=82, Z=50). The next highest yields are obtained for the fragment combinations $^{126}$Cd+$^4$He+$^{130}$Cd and $^{128}$Cd+$^4$He+$^{128}$Cd.

From the study of alpha accompanied ternary fission of even-even $^{244-260}$Cf isotopes, we have concluded that, the fragment combinations with the higher Q value and having doubly or near doubly magic nuclei as the heavier fragment possess the highest yield.

### 3.2. Effect of deformation and orientation of fragments

The effect of deformation and orientation of fragments in the alpha accompanied cold ternary fission of $^{244-260}$Cf isotopes have been analyzed by taking the interacting barrier as the sum of deformed Coulomb potential and deformed proximity potential. The Coulomb interaction between the two deformed and oriented nuclei, which is taken from [46] and which includes higher multipole deformation [47, 48], is given as,

$$V_C = \frac{Z_1 Z_2 e^2}{r} + 3Z_1 Z_2 e^2 \sum_{\lambda, i=1,2} \frac{1}{2\lambda+1} \frac{R_{0i}^\lambda}{r^{\lambda+1}} Y_\lambda^{(0)}(\alpha_i) \left[ \beta_{\lambda i} + \frac{4}{7} \beta_{\lambda i}^2 Y_\lambda^{(0)}(\alpha_i) \delta_{\lambda,2} \right] \quad (13)$$

with

$$R_i(\alpha_i) = R_{0i} \left[ 1 + \sum_\lambda \beta_{\lambda i} Y_\lambda^{(0)}(\alpha_i) \right] \quad (14)$$

where $R_{0i} = 1.28 A_i^{1/3} - 0.76 + 0.8 A_i^{-1/3}$. Here $\alpha_i$ is the angle between the radius vector and symmetry axis of the $i^{th}$ nuclei (see Fig.1 of Ref [47]) and it is to be noted that the quadrupole interaction term proportional to $\beta_{21}\beta_{22}$, is neglected because of its short range character.

In the case of proximity potential, $V_P(z) = 4\pi \gamma b \overline{R} \Phi(\varepsilon)$, the deformation comes only in the mean curvature radius. For spherical nuclei, mean curvature radius is defined as $\overline{R} = \frac{C_1 C_2}{C_1 + C_2}$, where $C_1$ and $C_2$ are Süssmann central radii of fragments. The mean curvature radius, $\overline{R}$ for two deformed nuclei lying in the same plane can be obtained by the relation,

$$\frac{1}{\overline{R}^2} = \frac{1}{R_{11} R_{12}} + \frac{1}{R_{21} R_{22}} + \frac{1}{R_{11} R_{22}} + \frac{1}{R_{21} R_{12}} \quad (15)$$

The four principal radii of curvature $R_{11}$, $R_{12}$, $R_{21}$ and $R_{22}$ are given by Baltz and Bayman [49].

With the inclusion of deformation and orientation of fragments, the driving potential is calculated for all fragment combinations found in the alpha accompanied ternary fission of even-even $^{244-260}$Cf isotopes. In the case of $^{244}$Cf isotope with $^4$He as light charged particle, the driving potential is plotted as a function of fragment mass number $A_1$ as shown in figure 2. The experimental values for quadrupole deformation ($\beta_2$) are taken from Ref. [50, 51] and for the cases where the experimental values are not available, we have taken those values from Moller et al [45]. In the cold reaction valley plot, three different cases are considered, (1) three fragments taken as spherical (2) two fragments ($A_1$ and $A_2$) as deformed with orientation $0^0 - 0^0$ and (3) two fragments ($A_1$ and $A_2$) as deformed with orientation $90^0 - 90^0$. It is clear from the plot that, in most of the cases, the fragment combinations with $0^0 - 0^0$ orientation have a low value for driving potential compared to the fragment combinations with $90^0 - 90^0$, but in few cases $90^0 - 90^0$ orientation has the low value. In $0^0 - 0^0$ orientation of fragments, both fragments are either prolate or one fragment is prolate and the other fragment is spherical, whereas in the case of $90^0 - 90^0$ orientation, both fragments are either oblate or one fragment is oblate and the other one is spherical. It is also noted that, the optimum fragment combinations change with the inclusion of deformation and orientation of fragments.

In fission process, the fragments are strongly polarized due to nuclear force and the fragments take either $0^0 - 0^0$ orientation or $90^0 - 90^0$ orientation. In the present manuscript we have considered only two orientations $0^0 - 0^0$ and $90^0 - 90^0$, because it is well established by Gupta and co-workers [48] that optimum orientation for prolate-prolate/spherical fragments is $0^0 - 0^0$, and for oblate-oblate/spherical fragments optimum orientation is $90^0 - 90^0$. It should be noted that

Manimaran et al [32] also considered only $0^0-0^0$ and $90^0-90^0$ orientations in their study on the effect of deformation and orientation in the $^4$He and $^{10}$Be accompanied ternary fission of $^{252}$Cf.

For example in the case of $^{244}$Cf isotope, the fragment combinations obtained with the spherical case $^{110}$Ru+$^4$He+$^{130}$Te and $^{120}$Cd+$^4$He+$^{120}$Cd changed to $^{110}$Nb+$^4$He+$^{130}$Cs and $^{120}$Ag+$^4$He+$^{120}$In respectively with the inclusion of deformation of fragments. In the case of $^{246}$Cf isotope, the fragment combinations $^{110}$Ru+$^4$He+$^{132}$Te and $^{120}$Cd+$^4$He+$^{122}$Cd changed to $^{110}$Zr+$^4$He+$^{132}$Ba and $^{120}$In+$^4$He+$^{122}$Ag respectively with the inclusion of deformation of fragments. For $^{248}$Cf isotope, the fragment combinations $^{104}$Mo+$^4$He+$^{140}$Xe and $^{120}$Cd+$^4$He+$^{124}$Cd obtained in spherical case changed to $^{104}$Ru+$^4$He+$^{140}$Te and $^{120}$Ag+$^4$He+$^{124}$In respectively with the inclusion of deformation. In the case of $^{250}$Cf isotope, the fragment combinations obtained in spherical case $^{108}$Mo+$^4$He+$^{138}$Xe and $^{112}$Ru+$^4$He+$^{134}$Te changed to $^{108}$Ru+$^4$He+$^{138}$Te and $^{112}$Zr+$^4$He+$^{134}$Ba respectively with the inclusion of deformation of fragments. For $^{252}$Cf isotope, the fragment combination $^{110}$Mo+$^4$He+$^{138}$Xe and $^{120}$Pd+$^4$He+$^{128}$Sn obtained in spherical case changed to $^{110}$Nb+$^4$He+$^{138}$Cs and $^{120}$Cd+$^4$He+$^{128}$Cd respectively with the inclusion of deformation. In the case of $^{254}$Cf isotope, $^{106}$Zr+$^4$He+$^{144}$Ba and $^{120}$Pd+$^4$He+$^{130}$Sn obtained in spherical case changed to $^{106}$Ru+$^4$He+$^{144}$Te and $^{120}$Ag+$^4$He+$^{130}$In respectively with the inclusion of deformation. For $^{256}$Cf isotope, the fragment combinations $^{104}$Zr+$^4$He+$^{148}$Ba and $^{116}$Ru+$^4$He+$^{136}$Te obtained with the spherical fragments changed to $^{104}$Tc+$^4$He+$^{148}$I and $^{116}$Pd+$^4$He+$^{136}$Sn respectively with the inclusion of deformation. In the case of $^{258}$Cf isotope, the fragment combination obtained with the spherical fragments $^{80}$Zn+$^4$He+$^{174}$Dy and $^{104}$Zr+$^4$He+$^{150}$Ba changed to $^{80}$Ga+$^4$He+$^{174}$Tb and $^{104}$Tc+$^4$He+$^{150}$I respectively with the inclusion of deformation of fragments. For $^{260}$Cf isotope, the fragment combination obtained with the spherical fragments $^{108}$Zr+$^4$He+$^{148}$Ba and $^{102}$Sr+$^4$He+$^{154}$Ce changed to $^{108}$Ru+$^4$He+$^{148}$Te and $^{102}$Mo+$^4$He+$^{154}$Xe respectively with the inclusion of deformation. For this reason we came to the conclusion that, the inclusion of deformation and orientation effects of the nuclei play a significant role in the alpha accompanied ternary fission of even-even $^{244-260}$Cf isotopes as that of closed shell effect.

The quadrupole deformation is included for all possible fragment combinations occurring in the alpha accompanied ternary fission of even-even $^{244-260}$Cf isotopes and hence the corresponding relative yield is calculated and plotted as a function of fragment mass number $A_1$ and $A_2$ as shown in figure 4(a) - 4(i). Here the calculations are done by taking the deformed Coulomb potential and deformed nuclear proximity potential. With the inclusion of quadrupole deformation ($\beta_2$), the width and height of the barrier are found to be reduced which in turn increases the barrier penetrability. For the alpha accompanied ternary fission of $^{244}$Cf and $^{246}$Cf isotopes, the highest yield is found for the fragment combination $^{114}$Pd+$^4$He+$^{126}$Sn and $^{114}$Pd+$^4$He+$^{128}$Sn respectively, both of these fragment combinations possess proton shell closure Z=50 of Sn nuclei. In the case of $^{248}$Cf and $^{252}$Cf isotopes, the highest yield is obtained for the fragment combinations $^{114}$Pd+$^4$He+$^{130}$Sn and $^{118}$Pd+$^4$He+$^{130}$Sn, in which $^{130}$Sn (N=80, Z=50) is a near doubly magic nuclei. For $^{250}$Cf, $^{254}$Cf, $^{256}$Cf, $^{258}$Cf and $^{260}$Cf isotopes, the highest yield is obtained for the fragment combination $^{114}$Pd+$^4$He+$^{132}$Sn, $^{118}$Pd+$^4$He+$^{132}$Sn, $^{120}$Pd+$^4$He+$^{132}$Sn, $^{122}$Pd+$^4$He+$^{132}$Sn and $^{124}$Pd+$^4$He+$^{132}$Sn respectively, all of which possess the presence of doubly magic nuclei $^{132}$Sn (N=82, Z=50).

In figure 5, the calculated yields obtained in the alpha accompanied ternary fission of $^{252}$Cf isotope are compared with the experimental data [15]. The calculations are made for the fragments considered as spherical and also for the fragments with the inclusion of quadrupole deformation ($\beta_2$).

From the graph it is clear that, the theoretical calculations we have made in the case of alpha accompanied ternary fission of $^{252}$Cf isotope are found to be agreement with the experimental data.

### 3.3. Emission probability of long range alpha particle

The alpha particle formed in the ternary fission process is fairly energetic with an average kinetic energy of 16MeV which is emitted in a direction perpendicular to the main fission fragments and hence termed as long range alpha particle. Carjan suggests that the emission of long range alpha particle is possible only if alpha particle is formed inside the fissioning nucleus so as to attain sufficient energy to overcome the Coulomb barrier of the scissioning nucleus. Based on the theoretical model of Carjan [20], Serot and Wagemans [21] suggested that the characteristics of long range alpha particle strongly depends on the preformation probability of alpha particle formed in the ternary fission process and also on various parameters like fissility parameter $Z^2/A$, alpha cluster preformation factor or spectroscopic factor $S_\alpha$ and fission modes.

The spectroscopic factor or alpha cluster preformation factor $S_\alpha$ can be calculated in a semi-empirical way which was proposed by Blendowske et al. [52] as, $S_\alpha = b\lambda_e / \lambda_{WKB}$, where $b$ is the branching ratio for the ground state to ground state transition, $\lambda_e$ is the experimental α decay constant and $\lambda_{WKB}$ is the α decay constant calculated from the WKB approximation.

The emission probability of long range alpha particle is determined with the number of fission events B and can be calculated as follows:

$$\frac{LRA}{B} = S_\alpha P_{LRA} \qquad (16)$$

where $S_\alpha$ is the alpha cluster preformation factor and $P_{LRA}$ is the probability of alpha particle preformed in the fissioning nucleus.

The probability of alpha particle preformed in the fissioning nucleus can be calculated as,

$$P_{LRA} = \exp\left\{-\frac{2}{\hbar}\int_{z_0}^{z_1}\sqrt{2\mu(V-Q)}dz\right\} \qquad (17)$$

The turning point $z_1 = 0$ represents the touching configuration and $z_0$ can be determined from the condition $V(z_0) = Q$, where $Q$ is the decay energy. For the internal (overlap) region, the potential is taken as a simple power law interpolation.

In ternary fission [53] the peak observed in the angular distribution of LRA in the direction perpendicular to the direction of main fission fragments, shows that LRA emission in ternary fission process is very similar to binary fission until the moment of scission. So the spectroscopic factor S and thereby the preformation factor $P_{LRA}$, which has been defined for two body problem in Ref [52] is suitable for ternary fission studies. It can be assumed that the LRA particle is emitted from the neck connecting the two fission fragments, most probably at the point at which the neck ruptures.

The emission probabilities of long range alpha particle in the ternary fission of even-even $^{244-260}$Cf isotopes are computed using the formalism discussed above and are given in table 1. The alpha cluster preformation factor $S_\alpha$ and corresponding probability for the alpha particle preformed in the fissioning nucleus $P_{LRA}$ are also listed in table 1. The calculated emission probabilities of long range alpha particle are found to agree well with the experimental data [23].

### 3.4. Kinetic energy of long range alpha particle

We have calculated the kinetic energy of long range alpha particle emitted in the ternary fission of $^{244-260}$Cf isotopes using the formalism reported by Fraenkel [54]. The conservation of total momentum in the direction of light fragment leads to the relation,

$$(m_L E_L)^{1/2} = (m_H E_H)^{1/2} \cos\theta_R - (m_\alpha E_\alpha)^{1/2} \cos\theta_L \qquad (18)$$

Here $m_L$, $m_H$ and $m_\alpha$ are the masses of the light, heavy and α particle respectively. $E_L$, $E_H$ and $E_\alpha$ represent the final energies of the light, heavy and α particle respectively. $\theta_L$ is the angle between the alpha particle and the light particle and $\theta_R$ is the recoil angle.

The conservation of total momentum in a direction perpendicular to light fragment leads to the relations for recoil angle $\theta_R$ as,

$$(m_H E_H)^{1/2} \sin\theta_R = (m_\alpha E_\alpha)^{1/2} \sin\theta_L \qquad (19)$$

Using eqn. (18) and eqn. (19), the kinetic energy of the long range alpha particle can be derived as follows,

$$E_\alpha = E_L \left(\frac{m_L}{m_\alpha}\right)(\sin\theta_L \cot\theta_R - \cos\theta_L)^{-2} \qquad (20)$$

In the present work in order to obtain the mean kinetic energy of alpha particle $E_\alpha$ = 16MeV and total fragment kinetic energy of the fissioning system as 168MeV, the recoil angle is taken as $\theta_R$ = 4.5° and angle between the alpha particle and the light particle is taken as $\theta_L$ = 92.25°. The kinetic energy of light fragment $E_L$ can be calculated as,

$$E_L = \frac{A_H}{A_L + A_H} TKE \qquad (21)$$

Here *TKE* is the total kinetic energy of fission fragments, which can be calculated using the expressions taken from Herbach et al. [55] as follows,

$$TKE = \frac{0.2904(Z_L + Z_H)^2}{A_L^{1/3} + A_H^{1/3} - (A_L + A_H)^{1/3}} \frac{A_L A_H}{(A_L + A_H)^2} \qquad (22)$$

where $A_L$ and $A_H$ are the mass numbers of light and heavy fragments respectively.

In the present manuscript, for the potential energy calculation, yield calculation and for the kinetic energy calculation we have assumed a triangular configuration. For computing kinetic energy [54] we have taken recoil angle $\theta_R$ as 4.5° for a 16MeV alpha particle emitted at an angle 90° with respect to the two main fragments. Further we would like to mention that one can use either triangular or collinear configuration for computing kinetic energy, as the size of alpha particle is small compared to the main fission fragments the initial phase for both the configurations are found to be the same.

The kinetic energy of long range alpha particle emitted in the ternary fission of $^{244-260}$Cf isotopes is calculated using the formalism described above and the corresponding experimental values are listed in table 2. The calculated results for which experimental data is available are in good agreement with each other [23].

## 4. Summary

The alpha accompanied cold ternary fission of even-even $^{244-260}$Cf isotopes has been studied using the Unified ternary fission model (UTFM). The relative yield is calculated by taking the interacting barrier as the sum of Coulomb potential and nuclear proximity potential. In the alpha accompanied cold ternary fission of $^{244}$Cf isotope, the highest yield is found for the fragment combination $^{108}$Ru+$^{4}$He+$^{132}$Te, in which $^{132}$Te (N=82, Z=52) is a near doubly magic nuclei. In the case of $^{246}$Cf and $^{248}$Cf isotopes, the highest yield is found for the fragment combination $^{108}$Ru+$^{4}$He+$^{134}$Te and $^{110}$Ru+$^{4}$He+$^{134}$Te, both of which possess near doubly magic nuclei $^{134}$Te (N=82, Z=52). For alpha accompanied ternary fission of $^{250}$Cf, $^{252}$Cf, $^{254}$Cf, $^{256}$Cf, $^{258}$Cf and $^{260}$Cf isotopes, the highest yield is obtained for the fragment combinations $^{114}$Pd+$^{4}$He+$^{132}$Sn, $^{116}$Pd+$^{4}$He+$^{132}$Sn, $^{118}$Pd+$^{4}$He+$^{132}$Sn, $^{120}$Pd+$^{4}$He+$^{132}$Sn, $^{122}$Pd+$^{4}$He+$^{132}$Sn and $^{124}$Pd+$^{4}$He+$^{132}$Sn respectively, all of which possess near doubly magic nuclei $^{132}$Sn (N=82, Z=50). With the inclusion of deformation and orientation of fragments into account, the driving potential and relative yield of all possible fragment combinations are studied in detail and found that, in addition to closed shell effect, ground state deformation also plays an important role in the alpha accompanied ternary fission of $^{244-260}$Cf isotopes. The computed isotopic yields for alpha accompanied ternary fission of $^{252}$Cf isotope are found to be agreement with the experimental data. The emission probability and kinetic energy of long range alpha particle is calculated for the various isotopes of Cf and the values obtained agree well with the experimental data.

**Acknowledgments**

The author KPS would like to thank the University Grants Commission, Govt. of India for the financial support under Major Research Project. No.42-760/2013 (SR) dated 22-03-2013.

**Table 1.** The calculated emission probability of long range alpha particle in the ternary fission of $^{244-260}$Cf isotopes and the corresponding experimental data [23] are listed. The corresponding spectroscopic factor $S_\alpha$ and $P_{LRA}$ are also listed.

| Isotope | $S_\alpha$ | $P_{LRA}$ | $\dfrac{LRA}{B}$ | $\left(\dfrac{LRA}{B}\right)_{EXP.}$ |
|---|---|---|---|---|
| $^{244}$Cf | 0.0195 | 0.1178 | 2.29 x 10$^{-3}$ | - |
| $^{246}$Cf | 0.0201 | 0.1365 | 2.74 x 10$^{-3}$ | - |
| $^{248}$Cf | 0.0165 | 0.1400 | 2.31 x 10$^{-3}$ | - |
| $^{250}$Cf | 0.0161 | 0.1631 | 2.63 x 10$^{-3}$ | (2.93 ± 0.10) x 10$^{-3}$ |
| $^{252}$Cf | 0.0252 | 0.2364 | 5.96 x 10$^{-3}$ | (2.56 ± 0.07) x 10$^{-3}$ |
| $^{254}$Cf | 0.0138 | 0.3109 | 4.29 x 10$^{-3}$ | - |
| $^{256}$Cf | 0.0128 | 0.3775 | 4.83 x 10$^{-3}$ | - |
| $^{258}$Cf | 0.0120 | 0.4591 | 5.51 x 10$^{-3}$ | - |
| $^{260}$Cf | 0.0112 | 0.5417 | 6.06 x 10$^{-3}$ | - |

**Table 2.** The calculated kinetic energy of alpha particle $E_\alpha$ emitted in the ternary fragmentation of $^{244-260}$Cf isotopes and the corresponding experimental data [23] are listed.

| Fragmentation channel | $E_\alpha$ (MeV) Calc. | $E_\alpha$ (MeV) Expt. | Fragmentation channel | $E_\alpha$ (MeV) Calc. | $E_\alpha$ (MeV) Expt. |
|---|---|---|---|---|---|
| $^{244}$Cf → $^{104}$Mo + $^4$He + $^{136}$Xe | 16.442 | | $^{252}$Cf → $^{120}$Pd + $^4$He + $^{128}$Sn | 17.295 | |
| $^{244}$Cf → $^{106}$Ru + $^4$He + $^{134}$Te | 16.560 | | $^{252}$Cf → $^{122}$Cd + $^4$He + $^{126}$Cd | 17.318 | |
| $^{244}$Cf → $^{108}$Ru + $^4$He + $^{132}$Te | 16.663 | | $^{252}$Cf → $^{124}$Cd + $^4$He + $^{124}$Cd | 17.325 | 15.96 ± 0.09 |
| $^{244}$Cf → $^{110}$Ru + $^4$He + $^{130}$Te | 16.751 | | $^{252}$Cf → $^{126}$Cd + $^4$He + $^{122}$Cd | 17.318 | |
| $^{244}$Cf → $^{112}$Pd + $^4$He + $^{128}$Sn | 16.823 | | $^{252}$Cf → $^{128}$Cd + $^4$He + $^{120}$Cd | 17.295 | |
| $^{246}$Cf → $^{108}$Ru + $^4$He + $^{134}$Te | 16.712 | | $^{254}$Cf → $^{116}$Ru + $^4$He + $^{134}$Te | 17.265 | |
| $^{246}$Cf → $^{110}$Ru + $^4$He + $^{132}$Te | 16.806 | | $^{254}$Cf → $^{118}$Pd + $^4$He + $^{132}$Sn | 17.325 | |
| $^{246}$Cf → $^{112}$Pd + $^4$He + $^{130}$Sn | 16.885 | | $^{254}$Cf → $^{120}$Pd + $^4$He + $^{130}$Sn | 17.371 | |
| $^{246}$Cf → $^{114}$Pd + $^4$He + $^{128}$Sn | 16.948 | | $^{254}$Cf → $^{122}$Pd + $^4$He + $^{128}$Sn | 17.401 | |
| $^{246}$Cf → $^{116}$Pd + $^4$He + $^{126}$Sn | 16.995 | | $^{254}$Cf → $^{124}$Cd + $^4$He + $^{126}$Cd | 17.416 | |
| $^{248}$Cf → $^{110}$Ru + $^4$He + $^{134}$Te | 16.858 | | $^{256}$Cf → $^{118}$Pd + $^4$He + $^{134}$Sn | 17.391 | |
| $^{248}$Cf → $^{112}$Pd + $^4$He + $^{132}$Sn | 16.943 | | $^{256}$Cf → $^{120}$Pd + $^4$He + $^{132}$Sn | 17.444 | |
| $^{248}$Cf → $^{114}$Pd + $^4$He + $^{130}$Sn | 17.013 | | $^{256}$Cf → $^{122}$Pd + $^4$He + $^{130}$Sn | 17.481 | |
| $^{248}$Cf → $^{116}$Pd + $^4$He + $^{128}$Sn | 17.068 | | $^{256}$Cf → $^{124}$Cd + $^4$He + $^{128}$Cd | 17.504 | |
| $^{248}$Cf → $^{118}$Pd + $^4$He + $^{126}$Sn | 17.107 | | $^{256}$Cf → $^{126}$Cd + $^4$He + $^{126}$Cd | 17.511 | |
| $^{250}$Cf → $^{112}$Ru + $^4$He + $^{134}$Te | 16.998 | | $^{258}$Cf → $^{118}$Ru + $^4$He + $^{136}$Te | 17.454 | |
| $^{250}$Cf → $^{114}$Pd + $^4$He + $^{132}$Sn | 17.075 | | $^{258}$Cf → $^{120}$Pd + $^4$He + $^{134}$Sn | 17.513 | |
| $^{250}$Cf → $^{116}$Pd + $^4$He + $^{130}$Sn | 17.137 | 15.95 ± 0.13 | $^{258}$Cf → $^{122}$Pd + $^4$He + $^{132}$Sn | 17.557 | |
| $^{250}$Cf → $^{118}$Pd + $^4$He + $^{128}$Sn | 17.184 | | $^{258}$Cf → $^{124}$Pd + $^4$He + $^{130}$Sn | 17.587 | |
| $^{250}$Cf → $^{120}$Pd + $^4$He + $^{126}$Sn | 17.215 | | $^{258}$Cf → $^{126}$Cd + $^4$He + $^{128}$Cd | 17.602 | |
| $^{252}$Cf → $^{110}$Mo + $^4$He + $^{138}$Xe | 16.952 | | $^{260}$Cf → $^{120}$Pd + $^4$He + $^{136}$Sn | 17.578 | |
| $^{252}$Cf → $^{112}$Ru + $^4$He + $^{136}$Te | 17.050 | | $^{260}$Cf → $^{122}$Pd + $^4$He + $^{134}$Sn | 17.629 | |
| $^{252}$Cf → $^{114}$Ru + $^4$He + $^{134}$Te | 17.134 | 15.96 ± 0.09 | $^{260}$Cf → $^{124}$Pd + $^4$He + $^{132}$Sn | 17.667 | |
| $^{252}$Cf → $^{116}$Pd + $^4$He + $^{132}$Sn | 17.203 | | $^{260}$Cf → $^{126}$Cd + $^4$He + $^{130}$Cd | 17.688 | |
| $^{252}$Cf → $^{118}$Pd + $^4$He + $^{130}$Sn | 17.256 | | $^{260}$Cf → $^{128}$Cd + $^4$He + $^{128}$Cd | 17.696 | |

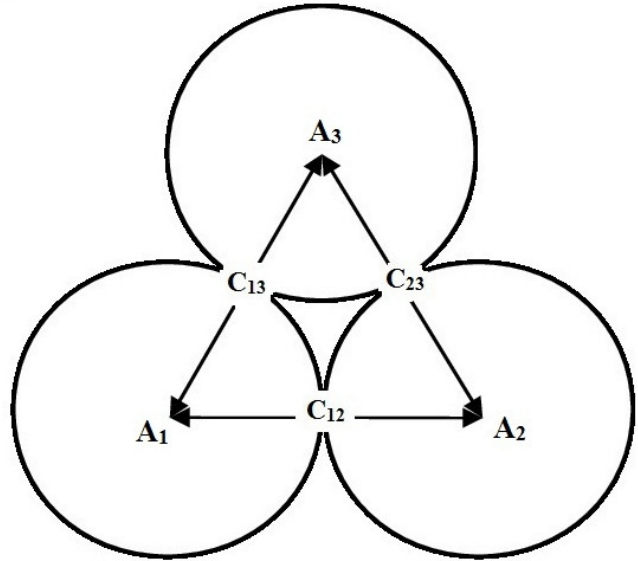

**Figure 1.** The touching configuration of three spherical fragments in equatorial configuration.

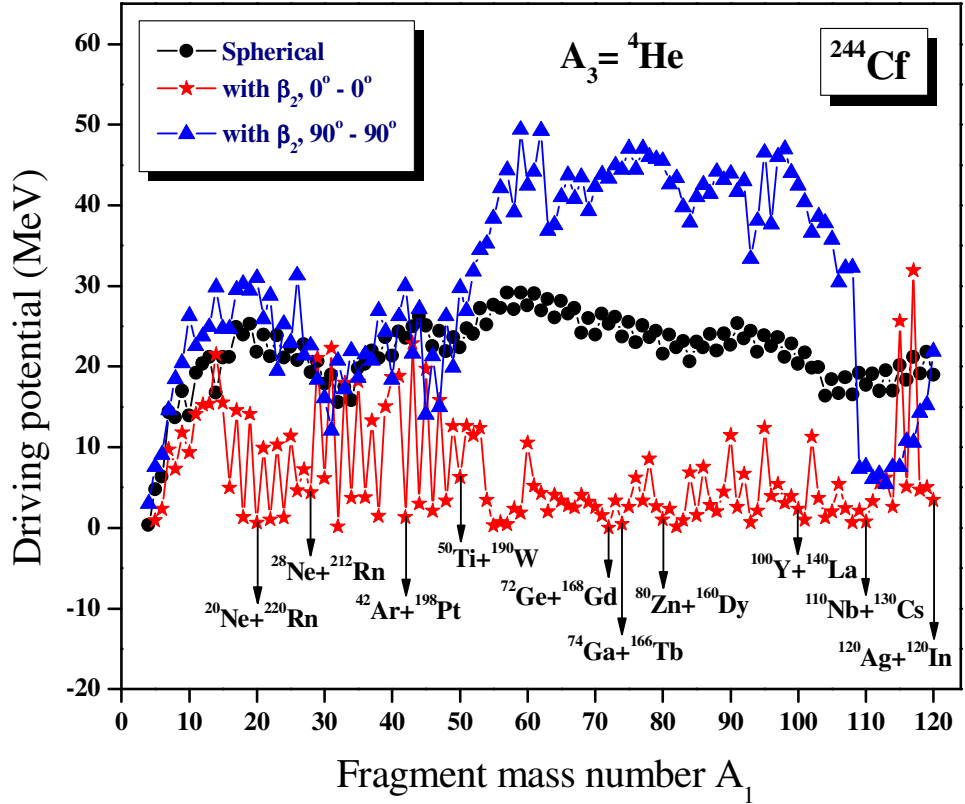

**Figure 2.** (Color online) The driving potential is plotted as a function of fragment mass number $A_1$ for the alpha accompanied ternary fission of $^{244}$Cf isotope with the inclusion of quadrupole deformation $\beta_2$ and for different orientation.

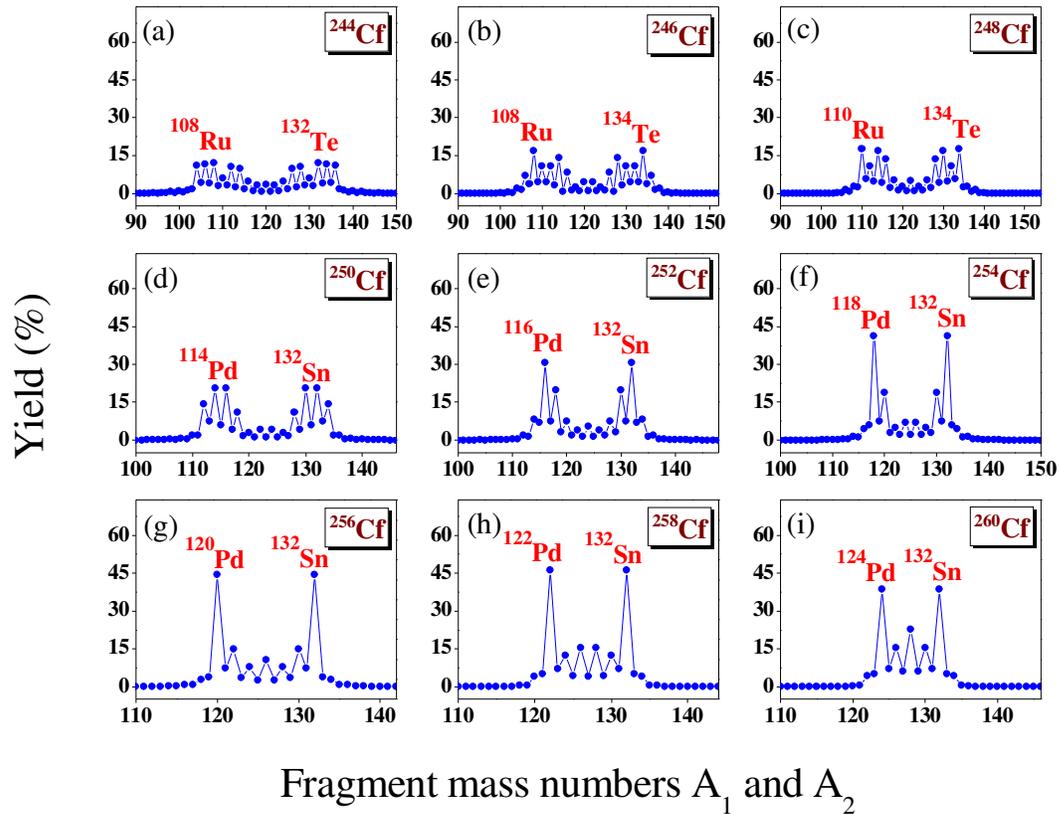

**Figure 3.** (Color online) The calculated yield is plotted as a function of fragment mass numbers $A_1$ and $A_2$ for the alpha accompanied cold ternary fission of even-even $^{244-260}$Cf isotopes with fragments treated as spherical.

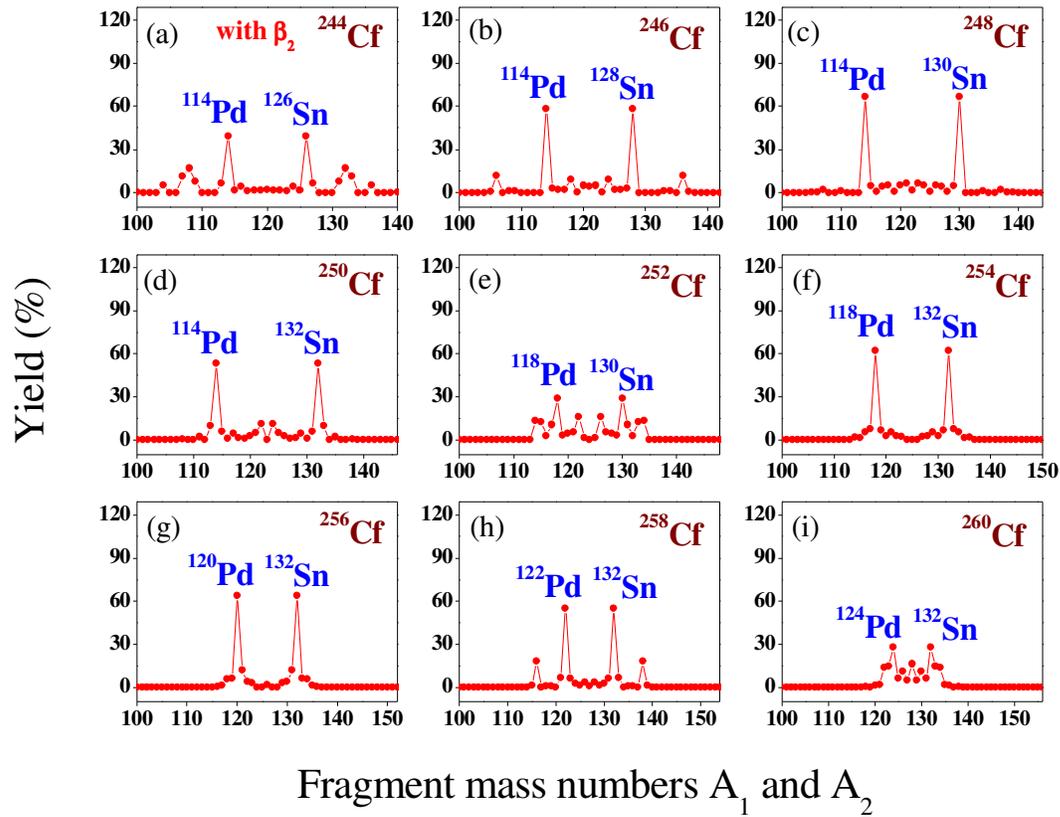

**Figure 4.** (Color online) The calculated yields for the alpha accompanied cold ternary fission of $^{244-260}$Cf isotopes plotted as a function of fragment mass numbers $A_1$ and $A_2$ with the inclusion of quadrupole deformation $\beta_2$.

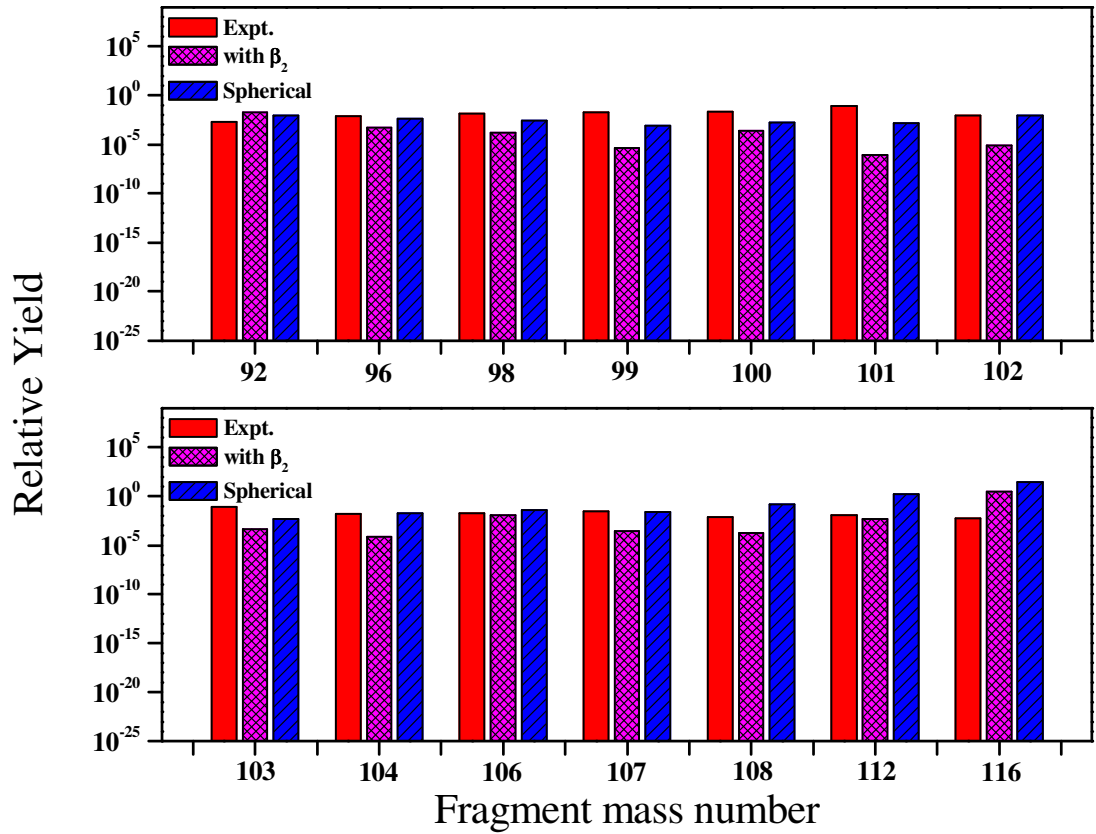

**Figure 5.** (Color online) The calculated yields obtained with the inclusion of quadrupole deformation of fragments and for the fragments considered as spherical are compared with the experimental data [15] in the alpha accompanied cold ternary fission of $^{252}$Cf isotope.